\documentclass[prb,aps,epsf,twocolumn,showpacs,superscriptaddress,10pt]{revtex4-1}
\usepackage{graphicx,amsfonts,times,bm,amsmath,verbatim,color,array}

\usepackage{subfigure}
\usepackage{float}
\usepackage{epsfig}
\usepackage{psfrag}
\usepackage{hyperref}
\hypersetup{
colorlinks=true,final=true,
        linkcolor=blue,
        citecolor=blue,
        filecolor=blue,
        urlcolor=blue,
}

\begin{document}

\title{Planar Thermal Hall Effect in Weyl Semimetals}

\author{S. Nandy}
\affiliation{Department of Physics, Indian Institute of Technology Kharagpur, W.B. 721302, India}
\author{A. Taraphder}
\affiliation{Department of Physics, Indian Institute of Technology Kharagpur, W.B. 721302, India}
\affiliation{Centre for Theoretical Studies, Indian Institute of Technology Kharagpur, W.B. 721302, India}
\affiliation{School of Basic Sciences, Indian Institute of Technology Mandi, Kamand 175005, India}
\author{Sumanta Tewari}
\affiliation{Department of Physics and Astronomy, Clemson University, Clemson, SC 29634,U.S.A}

\begin{abstract}
Weyl semimetals (WSMs) are intriguing topological states of matter that support various anomalous magneto-transport phenomena. One such phenomenon is a positive longitudinal magneto-conductivity and the associated planar Hall effect, which arise due to an effect known as chiral anomaly which is non-zero in the presence of electric and magnetic fields ($\mathbf{E} \neq 0$, $\mathbf{B} \neq 0$ and $\mathbf{E} \cdot \mathbf{B} \neq 0$). In this paper we show that another fascinating effect is the planar thermal Hall effect (PTHE), associated with positive longitudinal magneto-thermal conductivity (LMTC), which arise even in the absence of chiral anomaly ($\mathbf{E}= 0$, $\mathbf{B} \neq 0$). This effect is a result of chiral magnetic effect (CME) and involves the appearance of an in-plane transverse temperature gradient when the current due to a non-zero temperature gradiant ($\mathbf{\nabla} T$) and the magnetic field ($\mathbf{B}$) are not aligned with each other. Using semiclassical Boltzmann transport formalism in the relaxation time approximation we compute both longitudinal magneto-thermal conductivity and planar thermal Hall conductivity (PTHC) for a time reversal symmetry breaking WSM. We find that both LMTC and PTHC are quadratic in B in type-I WSM whereas each follows a linear-B dependence in type-II WSM in a configuration where $\mathbf{\nabla} T$ and B are applied along the tilt direction. In addition, we investigate the Wiedemann-Franz law for an inversion symmetry broken WSM (e.g., WTe$_{2}$) and find that this law is violated in these systems due to both chiral anomaly and CME.
\end{abstract}


\maketitle

\section{Introduction}
Dirac and Weyl semimetals (WSMs) have drawn tremendous attention of late due to their intriguing topological properties and anomalous response functions.
In these systems, the celebrated Dirac and Weyl equations, originally introduced for describing fundamental particles in high energy physics, become relevant for describing emergent, linearly dispersing, low energy excitations near gapless bulk nodes protected by topological invariants~\cite{Murakami_2007, Peskin_1995, Murakami2:2007, Yang:2011, Burkov1:2011, Burkov:2011, Volovik, Wan_2011, Xu:2011}. Weyl semimetals appear as topologically-nontrivial conductors where the spin-non-degenerate valence and conduction bands touch at isolated points in momentum space, the so called ``Weyl nodes". In WSMs, the Weyl nodes are separated in momentum space and always come in pairs of positive and negative monopole charges (also called chirality). The net monopole charge summed over all the Weyl points in the Brillouin zone vanishes~\cite{Nielsen:1981, Nielsen:1983}. The Weyl nodes act as the source and sink of Abelian Berry curvature, an analog of magnetic field but defined in the momentum space with quantized Berry flux~\cite{Xiao_2010}. In contrast to Dirac semimetals (DSMs) which are topologically protected in the presence of time reversal, space inversion, and additional spatial symmetries of the underlying crystal lattice, WSMs can be topologically protected \textit{in the absence} of time-reversal (TR) and/or space inversion (SI) symmetries~\cite{Volovik, Wan_2011, Yang:2011, Burkov1:2011, Burkov:2011, Zyuzin:2012, Xu:2011, Burkov_2012, Meng_2012, Gong_2011, Sau_2012, Hosur_2013} via the quantization of a topological invariant known as Chern number, defined as the non-zero quantized flux of the Berry curvature across any surface enclosing the bulk Weyl nodes. 

Several experimental groups have found evidence of the Weyl semimetal phase in inversion broken systems such as, TaAs~\cite{Lv_2015, Huang_2015, Hasan_2015}, WTe$_{2}$~\cite{Wu_2016}, MoTe$_{2}$~\cite{Jiang_2017}, and also in a 3D double gyroid photonic crystal~\cite{Lu_2015}, in the presence of TRS. There is another possible route to realize Weyl semimetal from a Dirac semimetal by breaking TRS externally using a magnetic field~\cite{Gorbar_2013}. The external magnetic field splits the Dirac cone of a DSM into a pair of Weyl cones even in the presence of inversion symmetry (IS). The TRS broken WSM contains a minimum of 2 Weyl nodes whereas the minimum number of Weyl nodes allowed in an inversion broken WSM is 4. For example, Bi$_{1-x}$Sb$_{x}$ for $x$ $\sim$ $3-4 \%$ is a Dirac semimetal~\cite{Fu_2007, Teo_2008, Guo_2011} which turns into a TR broken WSM in the presence of a magnetic field~\cite{Kim_2013}.

\begin{figure}[htb]
\begin{center}
\epsfig{file=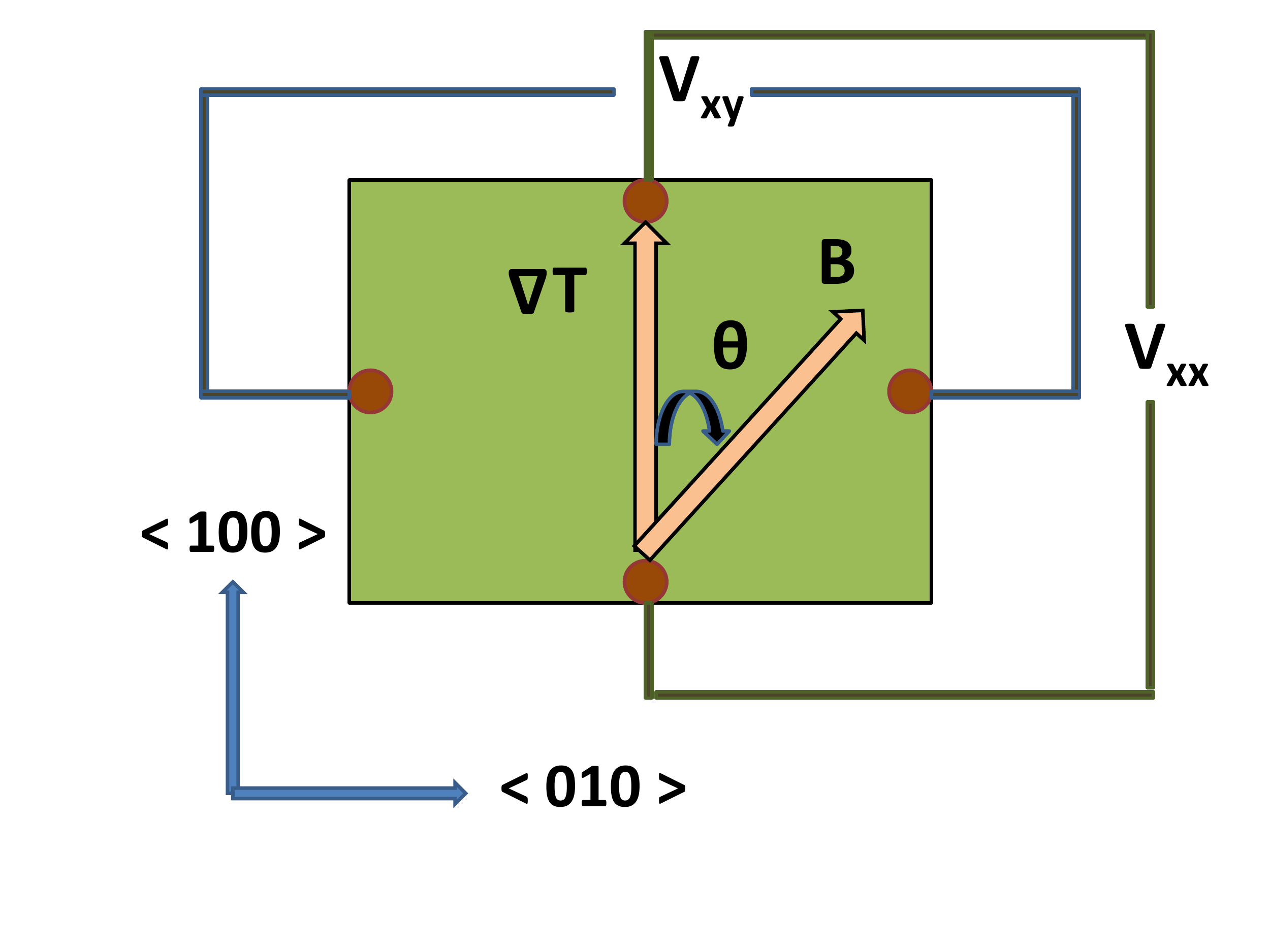,trim=0.10in 0.15in 0.0in 0.0in,clip=true, width=70mm}
\caption{(Color online) {Schematic illustration for the planar magneto-thermal transport measurement geometry. $<100>$ and $<010>$ denote the $x$ and $y$ directions respectively. Here, the temperature gradiant ($\nabla T$) is applied along the $<100>$ direction and the applied magnetic field is rotated in the $x-y$ plane by making an angle $\theta$ with the $x$ axis. $V_{xx}$ and $V_{xy}$ are the in-plane thermal gradiants across the $\nabla T$ and perpendicular to the $\nabla T$, respectively, which reveal the LMTC and PTHC in experiments.}}
\label{set-up_PTHC}
\end{center}
\end{figure}

From $\mathbf{k}\cdot\mathbf{p}$ theory, the low energy effective Hamiltonian near an isolated Weyl point situated at momentum space point $\mathbf{K}$ can be written as
\begin{eqnarray}
H_{\mathbf{k}} = \sum_{i=1}^{3} \; v_i (\mathbf{k}_i)\sigma_i,
\label{Eq_H_k_weyl_1}
\end{eqnarray}
where  $\hbar=c=1$, the crystal momenta $\mathbf{k}_i$ are measured from the band degeneracy point $\mathbf{K}$, and $\sigma_i$'s are the three Pauli matrices. The chirality of the Weyl point is defined by the sign of the product of the velocity components $\chi=\mathrm{sgn}(v_1 v_2 v_3)=\pm 1$. 
A fascinating transport signature due to non-trivial Berry curvature associated with Weyl nodes is the anomalous Hall effect in TR broken WSMs, where it depends linearly on the distance between the Weyl nodes in the momentum space~\cite{Burkov:2011}. In the presence of in-plane electric and magnetic fields, two other interesting topological effects, namely, negative longitudinal magneto-resistance (LMR) and planar Hall effect (PHE) appear due to non-conservation of separate electron numbers of opposite chirality for relativistic massless fermions, an effect known as the chiral or Adler-Bell-Jackiw anomaly~\cite{Goswami:2015,Zhong,Goswami:2013,Adler:1969, Bell:1969, Nielsen:1981, Nielsen:1983, Aji:2012, Zyuzin:2012, Volovik, Wan_2011, Xu:2011}. A number of theoretical~\cite{Kim:2014, Son:2013, Fiete_2014, Sharma:2016, Vladimir_2017, Burkov_jpcm, Pavan_2013, Burkov_2017, Nandy_2017, Nandy_2018} and experimental~\cite{Huang_2015, Jia_2016, Xu_2016, Erfu_2016, Li_2018, Liang_2018, Wang_2018, Chen_2018, Kumar_2018, Singha_2018} studies have been reported on chiral anomaly induced LMR and PHE. Replacing the electric field by a thermal gradient ($\mathbf{\nabla T}$), and for a parallel configuration between $\mathbf{\nabla T}$ and an applied field ($\mathbf{B}$), WSMs host another anomalous transport phenomenon known as positive longitudinal magneto-thermal conductivity (LMTC) which has recently been observed in experiments~\cite{Li_2016, Gooth_2017}. This effect arises from the so-called chiral magnetic effect (CME)~\cite{Kenji_2008, Franz_2013,Son_2012, Yin_2012, Chen_2013} - the generation of electric current along the direction of an external magnetic field triggered by chirality imbalance. The chiral electronics, an interesting application of the CME, refers to circuits with elements that have been proposed as quantum amplifiers of magnetic fields~\cite{Yee_2013}. In the present work, we propose another intriguing consequence of chiral magnetic effect in WSMs, the planar \textit{thermal-Hall} conductivity (PTHC), i.e., the appearance of an in-plane transverse temperature gradient ($V_{xy}$) when the co-planar $\mathbf{\nabla T}$ and $\mathbf{B}$ are not perfectly aligned to each other, precisely in a configuration in which the conventional and Berry-phase-mediated anomalous thermal Hall effect vanishes {as shown in Fig.~\ref{set-up_PTHC}}. 

In this paper we investigate the electronic contribution to LMTC and PTHC of type-I and type-II Weyl semimetals. It has been suggested earlier that the semiclassical Boltzmann equation approach is in good agreement with other theoretical approaches such as the Kubo formula and the quantum Boltzmann equation for thermal transport in WSMs~\cite{Burkov1:2011, Hosur_2012}. Furthermore, the Boltzmann equation gives exactly the same rate of change of the number of particles of a given chirality as relativistic quantum field theories~\cite{Son:2013}. Therefore, starting from the phenomenological Boltzmann transport equation in relaxation time approximation we derive the analytical expressions for LMTC and PTHC valid in the low field regime. In the present case, we study LMTC and PTHC for a lattice model of TRS broken WSM. We investigate the magnetic field dependence and angular dependence of LMTC and PTHC for two possible experimental set-ups. In the parallel set-up where both $\mathbf{\nabla T}$ and $\mathbf{B}$ are applied parallel to the tilt direction, LMTC and PTHC show different B-dependence and angular dependence for type-I and type-II WSMs. On the other hand LMTC and PTHC have similar dependence on B in the perpendicular set-up, i.e. with both $\mathbf{\nabla T}$ and $\mathbf{B}$ applied perpendicular to the tilt direction. 

The rest of the paper is organized as follows. In Sec.~\ref{model}, We introduce the lattice model of Weyl semimetal with broken TRS and explain the emergence of type-II WSM phase from type-I WSM phase. In Sec.~\ref{Boltzmann}, we solve the Boltzmann transport equation to obtain the analytical expression of PTHC in the presence of in-plane $E$ and $B$. In Sec.~\ref{Magneto}, we show our numerical results on LMTC and discuss the results in the context of two above mentioned possible experimental set-ups.  In Sec.~\ref{Hall}, we compute the PTHC for type-I and type-II Weyl semimetal. We discuss the magnetic field dependence and angular dependence of PTHC in both cases for two possible experimental set-ups. In Sec.~\ref{wiedemann}, we investigate the validity of Wiedemann-Franz law for an IS breaking type-II WSM WTe$_{2}$.
Finally in Sec.~\ref{summary}, we discuss the experimental aspects of the phenomena observed in our study and end with a conclusion.

\section{Model Hamiltonian}
\label{model}

\begin{figure*}[t]
\begin{center}
\epsfig{file=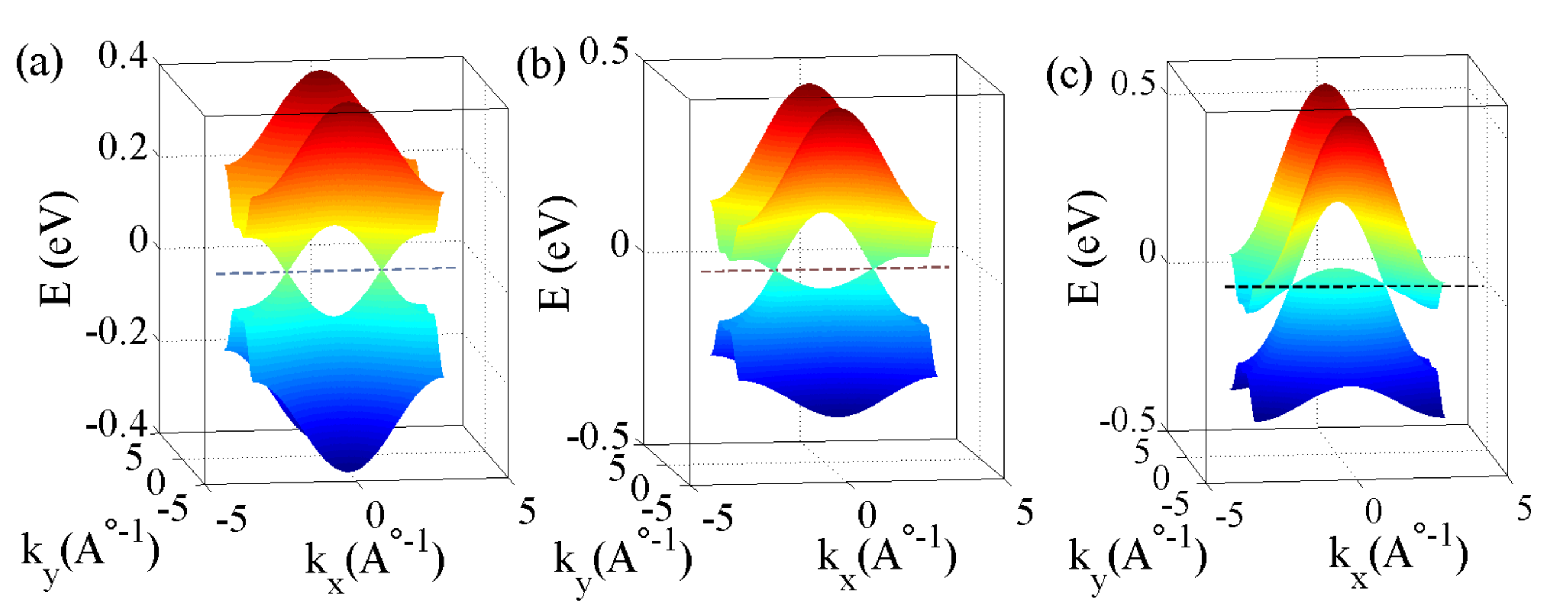,trim=0.0in 0.05in 0.0in 0.05in,clip=true, width=150mm}\vspace{0em}
\caption{(Color online) 3D band dispersions of the lattice model of Weyl fermions ($k_{z}$ is suppressed) obtained by diagonalizing Hamiltonian Eq.~(\ref{H_total}) for (a) $\gamma=0$, (b) $\gamma=0.05$ and (c) $\gamma=0.15$ respectively. The chemical potential is set at zero energy (indicated by dash line). The Weyl cones are at ($k_{0}$,0,0) and (-$k_{0}$,0,0). The parameters used are $t=-0.05$ eV, $m=0.15$ eV, and $k_{0}=\frac{\pi}{2}$.}
\label{lat_dis}
\end{center}
\end{figure*}

 We now discuss a prototype lattice model for a Weyl semimetal that breaks TRS but remains invariant under space inversion. Such a model, which possesses two Weyl nodes of opposite chirality tilted along $k_{x}$ direction, can be written as
\begin{equation}
{\cal H(\mathbf{k})}={H_{0}(\mathbf{k})}+{H_{T}(\mathbf{k})},
\label{H_total}
\end{equation}
where $H_{0}$ produces a pair of Weyl nodes of type-I at ($\pm k_{0}$,$0$,$0$)~\cite{Nandini_2017}.
\begin{eqnarray}
{H_{0}(\mathbf{k})}&=[m(\cos(k_{y}b)+\cos(k_{z}c)-2)+2t(\cos(k_{x}a)\nonumber \\
&-\cos k_{0})]\sigma_{1}-2t\sin(k_{y}b)\sigma_{2}-2t\sin(k_{z}c)\sigma_{3}. \nonumber \\
\label{H_lattice}
\end{eqnarray}
Here, $m$ is the mass and $t$ is hopping parameter.
The second term of the Hamiltonian  in Eq.~(2) which tilts  the nodes along $k_{x}$ direction can be written as,
\begin{equation}
{H_{T}(\mathbf{k})}=\gamma(\cos(k_{x}a)-\cos k_{0})\sigma_{0},
\label{H_tilt}
\end{equation}
where $\gamma$ is the tilt parameter which bends both the bands.
The 3D dispersion of the energy bands for different values of $\gamma$ are shown in Fig.~\ref{lat_dis}. When the anisotropy is zero (i.e $\gamma=0$), this Hamiltonian hosts nodes of type-I.  It is clear from Fig.~\ref{lat_dis}(b) that since the anisotropy along $k_{x}$ is small ($|\gamma|$ $<$ $|2t|$) the Fermi surface is still point-like; hence the Weyl nodes are still type-I. With further increase of the tilt parameter, a non-zero density of electron and hole states appear near the node energy for $\gamma$ $>$ $|2t|$. Thus, for $\gamma$ $>$ $|2t|$, the system is a type-II WSM as depicted in Fig.~\ref{lat_dis}(c). For this system, $\gamma$ $=$ $|2t|$ is the critical point between type-I and type-II WSM phases.
\section{Boltzmann Formalism For Planar Thermal Hall Conductivity}
\label{Boltzmann}
In this section, we focus on one specific response, namely, the planar thermal Hall effect that should be observed in all the Dirac and Weyl semimetals supporting negative longitudinal magneto-thermal conductivity. The planar thermal Hall effect is defined as an in-plane transverse temperature gradiant when the coplanar thermal gradiant and magnetic field are not perfectly aligned with each other.
Now we investigate the formulation of PTHC in the low field regime starting from the quasi-classical Boltzmann transport equation.

In the presence of an electric field ($\mathbf{E}$) and temperature gradiant ($\mathbf{\nabla T}$), the charge current ($\mathbf{J}$) and thermal current ($\mathbf{Q}$) can be written as
\begin{equation}
J_{\alpha}=L_{\alpha\beta}^{11}E_{\beta}+L_{\alpha\beta}^{12}(-\nabla_{\beta} T),
\label{e2}
\end{equation}
\begin{equation}
Q_{\alpha}=L_{\alpha\beta}^{21}E_{\beta}+L_{\alpha\beta}^{22}(-\nabla_{\beta} T),
\label{e3}
\end{equation}
where $\alpha$ and $\beta$ are spatial indices running over $x$, $y$, $z$, and $L$ represents different transport coefficients. In the presence of impurity scattering the phenomenological Boltzmann transport equation can be written as~\cite{John_2001}
\begin{equation}
\left(\frac{\partial}{\partial t}+\mathbf{\dot{r}} \cdot \mathbf{\nabla_{r}}+\mathbf{\dot{k}} \cdot \mathbf{\nabla_{k}}\right) f_\mathbf{k,r,t}=I_{coll} [f_{\mathbf{k,r,t}}],
\label{e5}
\end{equation}
where $f_\mathbf{k,r,t}$ is the electron distribution function. The right hand side $I_{coll}[f_{\mathbf{k,r,t}}]$ implies the collision integral incorporating electron correlations and impurity scattering effects. In the relaxation time approximation, the collision integral takes the following form $I_{coll}[f_{\mathbf{k}}]=\frac{f_{0}-f_{k}}{\tau(\mathbf{k})}$, where $\tau (\mathbf{k})$ is the intra-node relaxation time and $f_{0}$ is the equilibrium Fermi-Dirac distribution function in the absence of any external fields. {Here, we assume the inter-node scattering time ($\tau_s$) is much greater than intra-node scattering time ($\tau$), so that we only consider the intra-node scattering time in the current work.} We have also ignored momentum dependence of $\tau$ in the present work for simplifying the calculations. We treat intra-node scattering as a phenomenological parameter and assume $\tau^{+}=\tau^{-}$ for simplicity. Here, $\tau^{+}$ and $\tau^{-}$ are the scattering times appropriate for the two nodes. Dropping the r dependence of $f_\mathbf{k,r,t}$, valid for spatially uniform fields, and assuming steady state, the Boltzmann equation described by Eq.~(\ref{e5}) takes the following form
\begin{equation}
(\mathbf{\dot{r}} \cdot \mathbf{\nabla_{r}}+\mathbf{\dot{k}} \cdot \mathbf{\nabla_{k}})f_{k}=\frac{f_{0}-f_{k}}{\tau}.
\label{e6}
\end{equation}

It has been shown that in the presence of electric field and magnetic field, transport properties get substantially modified due to presence of non-trivial Berry curvature which acts as a fictitious magnetic field in the momentum space~\cite{Xiao_2010}. In addition to the band energy, the Berry curvature of the Bloch bands is required for a complete description of the electron dynamics in topological semimetals. The Berry curvature is defined by $\Omega(\mathbf{k})=\mathbf{\nabla_{k}} \times <v|i\mathbf{\nabla_{k}}|v>$ where $|v>$ is the periodic amplitude of the Bloch wave.

Using symmetry analysis, the general form of the Berry curvature can be obtained. Under time reversal symmetry, the Berry curvature follows $\mathbf{\Omega(-k)}=-\mathbf{\Omega(k)}$. On the other hand if the system has spatial inversion symmetry, then it follows that $\mathbf{\Omega(-k)}=\mathbf{\Omega(k)}$. Therefore, for a system with both time reversal and spatial inversion symmetries the Berry curvature vanishes identically throughout the Brillouin zone~\cite{Xiao_2010}. Conversely, if either time reversal symmetry or inversion symmetry is broken, the Berry curvature has non-zero values.

Incorporating the Berry curvature effects, the semi-classical equations of motion for an electron take the following form~\cite{Niu_1999, Niu_2006}
\begin{eqnarray}
\mathbf{\dot{r}}=\frac{1}{\hbar}\frac{\partial \epsilon_{\mathbf{k}}}{\partial \mathbf{k}}+ \frac{\mathbf{\dot{P}}}{\hbar} \times \mathbf{\Omega_{k}},
\label{e8a}
\end{eqnarray}
\begin{eqnarray}
\mathbf{\dot{P}}=e\mathbf{E}+e\mathbf{\dot{r}} \times \mathbf{B},
\label{e8b}
\end{eqnarray}
where the second term of Eq.~(\ref{e8a}) implies the anomalous velocity due to $\mathbf{\Omega_{k}}$. The Berry curvature carries an opposite sign for Weyl nodes of opposite chirality.
After solving two coupled equations for $\mathbf{\dot{r}}$ and $\mathbf{\dot{P}}$, we obtain the following modified semiclassical equations of motion~\cite{Duval_2006, Son_2012}
\begin{align}
\mathbf{\dot{r}}=\frac{1}{D(\mathbf{B,\Omega_{k}})}[\mathbf{v_{k}}+\frac{e}{\hbar}(\mathbf{E}\times\mathbf{\Omega_{k}})+\frac{e}{\hbar}(\mathbf{v_{k}}\cdot\mathbf{\Omega_{k}})\mathbf{B}],
\label{e9}
\end{align}
\begin{align}
\hbar\mathbf{\dot{k}}=\frac{1}{D(\mathbf{B,\Omega_{k}})}[e\mathbf{E}+\frac{e}{\hbar}(\mathbf{v_{k}} \times \mathbf{B})+\frac{e^{2}}{\hbar}(\mathbf{E}\cdot\mathbf{B})\mathbf{\Omega_{k}}],
\label{e10}
\end{align}

where $D(\mathbf{B,\Omega_{k}})=\left[1+\frac{e}{\hbar}(\mathbf{B} \cdot \mathbf{\Omega_{k}})\right]$, and $\mathbf{v_{k}}=\frac{1}{\hbar}\frac{\partial \epsilon_{\mathbf{k}}}{\partial \mathbf{k}}$ is the group velocity. The factor $D(\mathbf{B,\Omega_{k}})$ modifies the invariant phase space volume according to $dkdx \rightarrow D^{-1}(\mathbf{B,\Omega_{k}})dkdx$, giving rise to a noncommutative mechanical model, because the Poisson brackets of these co-ordinates is non-zero~\cite{Duval_2006}. So from hereon, we use $D=D(\mathbf{B,\Omega_{k}})$ in the rest of the paper for  simplicity.

The third term in Eq.~(\ref{e9}) gives rise to chiral magnetic effect. The chiral magnetic effect, an interesting signature of transport phenomena in Weyl semimetals, appears for $\mathbf{E}=0$~\cite{Son_2012, Yin_2012, Chen_2013, Kenji_2008}. It has been shown that electric currents ($\propto \mathbf{B}$) flow along the direction of the magnetic field in Weyl semimetals without any electric field in the presence of finite chiral chemical potential ($\mu_{+}-\mu_{-}$) where $\mu_{+}$ and $\mu_{-}$ imply the chemical potentials of the two Weyl nodes respectively~\cite{Kim:2014}. It has been discussed that the chiral magnetic effect depends on the limiting procedure for the transferred momentum and frequency~\cite{Pavan_2013}. In the dc limit i.e. when the frequency is set to zero first, the system is in equilibrium and the chiral magnetic effect vanishes. On the other hand, when the momentum $q$ is set to zero first, the system is away from equilibrium and the chiral magnetic effect does not vanish~\cite{Chen_2013}. The second term on the right hand side of the Eq.~(\ref{e10}) gives the usual Lorentz force, and the third term arises from chiral anomaly.

In order to compute the PTHC, we applied a temperature gradiant ($\nabla T$) along the $x$ axis and the magnetic field (B) is rotated in the $x-y$ plane in the absence of electric field i.e. $\mathbf{B}=B\cos\theta\hat{x}+B\sin\theta\hat{y}$, $\mathbf{\nabla T}=\nabla T\hat{x}$, $\mathbf{E}=0$. Here, $\theta$ is the angle between applied $\nabla T$ and $\mathbf{B}$ {as shown in Fig.~\ref{set-up_PTHC}}.
After substituting $\mathbf{\dot{r}}$ and $\mathbf{\dot{k}}$ described in Eq.~(\ref{e9}) and Eq.~(\ref{e10}) into Eq.~(\ref{e6}), the quasi-classical Boltzmann equation takes the following form
\begin{align}
&\left[\mathbf{v_{k}}+\frac{e}{\hbar}(\mathbf{v_{k}}\cdot \mathbf{\Omega_{k}})\mathbf{B}\right]\cdot \mathbf{\nabla_{r}}f_{\mathbf{k}}+\frac{eB}{\hbar^{2}}[(v_{x}\sin\theta-v_{y}\cos\theta)\frac{\partial}{\partial k_{z}}  \nonumber \\
&+v_{z}\cos\theta\frac{\partial}{\partial k_{y}}-v_{z}\sin\theta\frac{\partial}{\partial k_{x}}]f_{\mathbf{k}}=\frac{D(f_{0}-f_{\mathbf{k}})}{\tau}.
\label{e20}
\end{align}
Using the relation $\frac{\partial f_{0}}{\partial T}$ $=$ $\frac{(\epsilon-\mu)}{T}(-\frac{\partial f_{0}}{\partial \epsilon})$ ($\mu$ is the chemical potential) and assuming linear response, the above equation becomes
\begin{align}
&\frac{(\epsilon-\mu)\nabla T}{DT}\left[{v_{x}}+\frac{eB \cos \theta}{\hbar}(\mathbf{v_{k}}\cdot \mathbf{\Omega_{k}})\right]\left(-\frac{\partial f_{0}}{\partial \epsilon}\right)+\frac{eB}{\hbar^{2}}[(v_{x}\sin\theta \nonumber \\
&-v_{y}\cos\theta)\frac{\partial}{\partial k_{z}}
+v_{z}\cos\theta\frac{\partial}{\partial k_{y}}-v_{z}\sin\theta\frac{\partial}{\partial k_{x}}]f_{\mathbf{k}}=\frac{(f_{0}-f_{\mathbf{k}})}{\tau}.
\label{e21}
\end{align}
Now we attempt to solve the above equation by assuming the following ansatz for the electron distribution function deviation $\delta f_{\mathbf{k}}=f_{\mathbf{k}}-f_{0}$
\begin{align}
\delta f_{\mathbf{k}}= \frac{\tau(\epsilon-\mu)}{D}\frac{\nabla T}{T}\left[{v_{x}}+\frac{eB \cos \theta}{\hbar}(\mathbf{v_{k}} \cdot \mathbf{\Omega_{k}})-\mathbf{v} \cdot \mathbf{\zeta}\right]\left(\frac{\partial f_{0}} {\partial \epsilon}\right),
\label{e22}
\end{align}
where $\mathbf{\zeta}$ is the correction factor to account magnetic field. Plugging $f_{\mathbf{k}}$ into Eq.~(\ref{e21}), we have
\begin{align}
&\frac{eB}{\hbar^{2}}\left[(v_{x}\sin\theta-v_{y}\cos\theta)\frac{\partial}{\partial k_{z}}
+v_{z}\cos\theta\frac{\partial}{\partial k_{y}}-v_{z}\sin\theta\frac{\partial}{\partial k_{x}}\right] \nonumber \\
&\left[\frac{\tau(\epsilon-\mu)}{D}\frac{\nabla T}{T}({v_{x}}+\frac{eB \cos \theta}{\hbar}(\mathbf{v_{k}}\cdot\mathbf{\Omega_{k}}))-\mathbf{v}.
\mathbf{\zeta}\right]=\frac{D(\mathbf{v}\cdot\mathbf{\zeta})}{\tau}.
\label{e23}
\end{align}
We will now calculate the correction factor $\mathbf{\zeta}$ which vanishes in the absence of magnetic field B. This can be evaluated by expanding the inverse band-mass tensor which arises in Eq.~(\ref{e23}), and noting the fact that the above equation is valid for all values of velocity. Substituting the expression of band-mass tensor $m_{\alpha \beta}^{*}=\frac{1}{\hbar^{2}}\frac{\partial^{2} \epsilon_{\mathbf{k}}}{\partial k_{\alpha}\partial k_{\beta}}$, the above equation takes the following form
\begin{align}
&eB\frac{(\epsilon-\mu)}{DT}\nabla T[v_{z}(\frac{\sin \theta}{m_{xx}^{*}}-\frac{\cos \theta}{m_{xy}^{*}})-\frac{(v_{x}\sin\theta-v_{y}\cos\theta)}{m_{xz}^{*}}+ \nonumber \\
&\frac{eB\cos \theta}{\hbar}(v_{z}C_{1}\sin \theta-v_{z}C_{2}\cos \theta-(v_{x}\sin\theta-v_{y}\cos\theta)C_{3})] \nonumber \\
&+eB[-v_{z}\sin \theta(\frac{\zeta_{x}}{m_{xx}^{*}}+\frac{\zeta_{y}}{m_{xy}^{*}}+\frac{\zeta_{z}}{m_{xz}^{*}})+v_{z}(\frac{\zeta_{x}}{m_{xy}^{*}}+\frac{\zeta_{y}}{m_{yy}^{*}} \nonumber \\
&+\frac{\zeta_{z}}{m_{yz}^{*}})\cos \theta+(v_{x}\sin\theta-v_{y}\cos\theta)(\frac{\zeta_{x}}{m_{xz}^{*}}+\frac{\zeta_{y}}{m_{yz}^{*}}+\frac{\zeta_{z}}{m_{zz}^{*}})] \nonumber \\
&=-\frac{D}{\tau}(v_{x}\zeta_{x}+v_{y}\zeta_{y}+v_{z}\zeta_{z}),
\label{e24}
\end{align}
where we have identified $C_{1}$, $C_{2}$, and $C_{3}$ as
\begin{align}
&C_{1}=\frac{\Omega_{x}}{m_{xx}^{*}}+\frac{\Omega_{y}}{m_{xy}^{*}}+\frac{\Omega_{z}}{m_{xz}^{*}},\nonumber \\
&C_{2}=\frac{\Omega_{x}}{m_{xy}^{*}}+\frac{\Omega_{y}}{m_{yy}^{*}}+\frac{\Omega_{z}}{m_{yz}^{*}}; C_{3}=\frac{\Omega_{x}}{m_{xz}^{*}}+\frac{\Omega_{y}}{m_{yz}^{*}}+\frac{\Omega_{z}}{m_{zz}^{*}}.
\label{e25}
\end{align}
Now imposing the condition that the Eq.~(\ref{e24}) is valid for all values of $v_{x}$, $v_{y}$, and $v_{z}$, the correction factors $\zeta_{x}$, $\zeta_{y}$, and $\zeta_{z}$ can be calculated by evaluating the equation. After some straightforward algebra, we can write down the correction factors as given below.
\begin{align}
&\zeta_{z}=\frac{N_{0}(\alpha_{1}\alpha_{2}-\alpha_{3}\alpha_{4})}{\frac{D^{2}}{\tau^{2}}-(\frac{eB\cos \theta}{m_{yz}}-\frac{eB\sin \theta}{m_{xz}^{*}})^{2}-\frac{eB}{m_{zz}^{*}}\alpha_{4}}, \nonumber \\
&\zeta_{y}=\frac{\cos \theta\left[N_{0}(\frac{1}{m_{xz}^{*}}+\frac{eBC_{3}\cos \theta}{\hbar})-\zeta_{z}\frac{eB}{m_{zz}^{*}}\right]}{(\frac{eB\cos \theta}{m_{yz}^{*}}-\frac{D}{\tau}-\frac{eB\sin \theta}{m_{xz}^{*}})}, \nonumber \\
&\zeta_{x}=-\tan \theta \zeta_{y},
\label{e26}
\end{align}
where $N_{0}$, $\alpha_{1}$, $\alpha_{2}$, $\alpha_{3}$ and $\alpha_{4}$ can be written as
\begin{align}
&N_{0}=eB\tau \nabla T \frac{(\epsilon_{k}-\mu)}{DT}, \nonumber \\
&\alpha_{1}=\frac{\sin \theta}{m_{xx}^{*}}-\frac{\cos \theta}{m_{xy}^{*}}+\frac{eB\cos \theta}{\hbar}(C_{1}\sin \theta-C_{2}\cos \theta) \nonumber \\
&\alpha_{2}=\frac{eB\cos \theta}{m_{yz}^{*}}-\frac{D}{\tau}-\frac{eB\sin \theta}{m_{xz}^{*}}; \alpha_{3}=\frac{eB\cos \theta}{\hbar}C_{3}+\frac{1}{m_{xz}},\nonumber \\
&\alpha_{4}=\frac{eB\sin 2\theta}{m_{xy}^{*}}-\frac{eB\cos^{2} \theta}{m_{yy}^{*}}-\frac{eB\sin^{2} \theta}{m_{xx}^{*}}. \nonumber \\
\label{e27}
\end{align}
With all the correction factors in hand, we can now write the Boltzmann distribution function $f_{k}$ explicitly by using the Eq.~(\ref{e22})
\begin{align}
f_{\mathbf{k}}&=f_{0}+\tau \nabla T \frac{(\epsilon_{k}-\mu)}{DT}\left[{v_{x}}+\frac{eB\cos \theta}{\hbar}(\mathbf{v_{k}}.\mathbf{\Omega_{k}})\right]\left(\frac{\partial f_{0}}{\partial\epsilon}\right) \nonumber \\
&-\tau \nabla T \frac{(\epsilon_{k}-\mu)}{DT}(v_{x}d_{x}\sin \theta+v_{y}d_{y}\cos \theta+v_{z}d_{z})\left(\frac{\partial f_{0}}{\partial\epsilon}\right), \nonumber \\
\label{e28}
\end{align}
where $d_{x}$, $d_{y}$, and $d_{z}$, incorporating Berry phase effects, are related to correction factors $\zeta$ by the following relation.
\begin{align}
&\zeta_{x}=\tau \nabla T \frac{(\epsilon_{k}-\mu)}{DT} d_{x}\sin \theta, \nonumber \\
&\zeta_{y}=\tau \nabla T \frac{(\epsilon_{k}-\mu)}{DT} d_{y}\cos \theta; \zeta_{z}=\tau \nabla T \frac{(\epsilon_{k}-\mu)}{DT} d_{z}.
\label{e29}
\end{align}

Now in the presence of thermal gradiant and applied magnetic field, the thermal current takes the following form after accounting for both normal and anomalous contributions~\cite{Niu_2006, Shi_2011, Bergman_2010, Murakami_2011}
\begin{align}
\mathbf{Q}&=\int\frac{d^{3}k}{(2\pi)^{3}}(\epsilon_{\mathbf{k}}-\mu)D^{-1}\left[{\mathbf{v_{k}}}+\frac{e\mathbf{B}}{\hbar}(\mathbf{v_{k}}\cdot\mathbf{\Omega_{k}})\right]f_{\mathbf{k}}+\frac{k_{B}\mathbf{\nabla T}}{\beta \hbar}\nonumber \\
&\times \int \frac{d^{3}k}{(2\pi)^{3}}\mathbf{\Omega_{k}}\left[\frac{\pi^{2}}{3}f_{0}+\beta^{2}(\epsilon-\mu)^{2}f_{0}\right]-\frac{k_{B}\mathbf{\nabla T}}{\beta \hbar}\nonumber \\
&\times \int \frac{d^{3}k}{(2\pi)^{3}}\mathbf{\Omega_{k}}\left[ln^{2}(1+e^{-\beta(\epsilon_{\mathbf{k}}-\mu)^{2}})+2Li_{2}(1-f_{0})\right],
\label{e18}
\end{align}
where the first term of the above equation represents the standard contribution to the heat current in the absence of Berry curvature. Here, Li$_{2}$(z) is the polylogarithmic function of order 2, defined as
\begin{equation}
Li_{n}(z)=\sum^{\infty}_{k=1} \frac{z^{k}}{k^{n}}
\end{equation}
for an arbitrary complex order n, for a complex argument $|z| < 1$. The other terms of Eq.~(\ref{e18}) implies the anomalous response of the heat current. In the application of thermal gradiant, the anomalous response of Q can be written as $Q_{x}=l_{xy}\nabla_{y} T$. The quantity $l_{xy}$ can be calculated using the relation $l_{xy}=-\frac{k^{2}_{B}Tc_{2}}{\hbar}$ where $c_{2}$ can be written as~\cite{Bergman_2010}
\begin{equation}
c_{m}=\int [dk]\Omega_{z} \int^{\infty}_{\epsilon-\mu} d\epsilon (\beta \epsilon)^{m} \frac{\partial f_{0}}{\partial \epsilon}.
\label{e60}
\end{equation}
For $m=2$, the energy integral described in Eq.~(\ref{e60}) reduces to following form~\cite{Bergman_2010, Murakami_2011}
\begin{align}
\int^{\infty}_{\epsilon-\mu} d\epsilon (\beta \epsilon)^{2} \frac{\partial f_{0}}{\partial \epsilon}&=\frac{\pi^{2}}{3}f_{0}+\beta^{2}(\epsilon-\mu)^{2}f_{0} \nonumber \\
&-ln^{2}(1+e^{-\beta(\epsilon_{\mathbf{k}}-\mu)^{2}})-2Li_{2}(1-f_{0}).
\end{align}

Substituting $f_{k}$ in Eq.~(\ref{e18}) and comparing with the linear response Eq.~(\ref{e3}), we now arrive at the expression for longitudinal magneto-thermal conductivity
\begin{align}
l_{xx}=&\int\frac{d^{3}k}{(2\pi)^{3}}D^{-1}\tau \frac{(\epsilon_{\mathbf{k}}-\mu)^{2}}{T}[({v_{x}}+\frac{eB\cos \theta}{\hbar}(\mathbf{v_{k}}\cdot \mathbf{\Omega_{k}}))^{2} \nonumber \\
&-(\sin \theta d_{x}v_{x}+\cos \theta d_{y}v_{y}+d_{z}v_{z})({v_{x}}+\frac{eB\cos \theta}{\hbar}\nonumber \\
&(\mathbf{v_{k}}\cdot\mathbf{\Omega_{k}}))]\left(-\frac{\partial f_{0}}{\partial \epsilon}\right)=L_{xx}^{22}.
\label{e30}
\end{align}
In the limit of $\theta=0$, we get back to the same equation as discussed in earlier works~\cite{Kim:2014, Fiete_2014, Sharma:2016, Vladimir_2017}. In similar way, we can write down the thermal Hall conductivity as
\begin{align}
l_{yx}&=\int\frac{d^{3}k}{(2\pi)^{3}}\tau\frac{(\epsilon_{\mathbf{k}}-\mu)^{2}}{DT}\left(\frac{\partial f_{0}}{\partial \epsilon}\right)\{\left[v_{y}+\frac{eB\sin \theta}{\hbar}(\mathbf{v_{k}}\cdot\mathbf{\Omega_{k}})\right] \nonumber \\
&\left[v_{x}+\frac{eB\cos \theta}{\hbar}(\mathbf{v_{k}}\cdot \mathbf{\Omega_{k}})\right]\}+ \int \frac{d^{3}k}{(2\pi)^{3}}\tau\frac{(\epsilon_{\mathbf{k}}-\mu)^{2}}{T}[\sin \theta d_{x}v_{x} \nonumber \\
&+\cos \theta d_{y}v_{y}+d_{z}v_{z}]\left[v_{y}+ \frac{eB\sin \theta}{\hbar}(\mathbf{v_{k}}\cdot \mathbf{\Omega_{k}}))\right]\left(-\frac{\partial f_{0}}{\partial \epsilon}\right) \nonumber \\
&+\frac{k_{B}}{\beta \hbar} \int \frac{d^{3}k}{(2\pi)^{3}}\Omega_{z}\left[\ln^{2}(1+e^{-\beta(\epsilon_{k}-\mu)^{2}})+2Li_{2}(1-f_{0})\right] \nonumber \\
&-\frac{k_{B}}{\beta \hbar} \int \frac{d^{3}k}{(2\pi)^{3}} f_{0} \Omega_{z}\left[\frac{\pi^{2}}{3}+\beta^{2}(\epsilon_{k}-\mu)^{2}\right]=L_{yx}^{22}. \nonumber \\
\label{e31}
\end{align}
In the present work we are only interested in the chiral magnetic effect induced contribution to the planar thermal Hall conductivity. Therefore, we do not consider the last two terms of the above equation any further because these terms leads to the Berry curvature induced anomalous thermal Hall contribution in the absence of magnetic field which vanish in inversion symmetry breaking WSM. Neglecting the terms which are of a much smaller order of magnitude compared to others in typical Weyl metals, the final simplified expression of chiral magnetic effect induced longitudinal magneto-thermal conductivity ($l_{xx}$) and planar thermal Hall conductivity ($l_{yx}^{\text{pth}}$) can be written as
\begin{eqnarray}
l_{xx}=-\int\frac{d^{3}k}{(2\pi)^{3}}\tau \frac{(\epsilon_{\mathbf{k}}-\mu)^{2}}{DT}\left[{v_{x}}+\frac{eB\cos \theta}{\hbar}(\mathbf{v_{k}} \cdot\mathbf{\Omega_{k}})\right]^{2}\frac{\partial f_{0}}{\partial \epsilon}, \nonumber \\
\label{e32}
\end{eqnarray}
\begin{align}
l_{yx}^{\text{pth}}&=\int\frac{d^{3}k}{(2\pi)^{3}}D^{-1}\tau\frac{(\epsilon_{\mathbf{k}}-\mu)^{2}}{T}\left(-\frac{\partial f_{0}}{\partial \epsilon}\right) [\frac{eB\sin \theta}{\hbar}(\mathbf{v_{k}}\cdot\mathbf{\Omega_{k}}) \nonumber \\
& (v_{x}+\frac{eB\cos \theta}{\hbar}(\mathbf{v_{k}}\cdot\mathbf{\Omega_{k}}))].
\label{e33}
\end{align}
{Similarly, when the temperature gradiant is along the $z$ axis and the magnetic field is rotated in the $z-y$ plane, the expression of the LMTC takes the form 
\begin{eqnarray}
l_{zz}=-\int\frac{d^{3}k}{(2\pi)^{3}}\tau \frac{(\epsilon_{\mathbf{k}}-\mu)^{2}}{DT}\left[{v_{z}}+\frac{eB\cos \theta}{\hbar}(\mathbf{v_{k}} \cdot\mathbf{\Omega_{k}})\right]^{2}\frac{\partial f_{0}}{\partial \epsilon}. \nonumber \\
\label{e34}
\end{eqnarray}}
{It is clear from Eq.~(\ref{e32}), Eq.~(\ref{e33}) and Eq.~(\ref{e34}) that both LMTC and planar thermal Hall conductivity (PTHC) are Fermi-Surface quantities. In this work, we set the chemical potential $\mu > 0$. Therefore we use Eqs. (29, 31) only with respect to the conduction band to calculate these quantities.}
\section{Longitudinal Magneto-thermal conductivity}
\label{Magneto}
\begin{figure}[htb]
\begin{center}
\epsfig{file=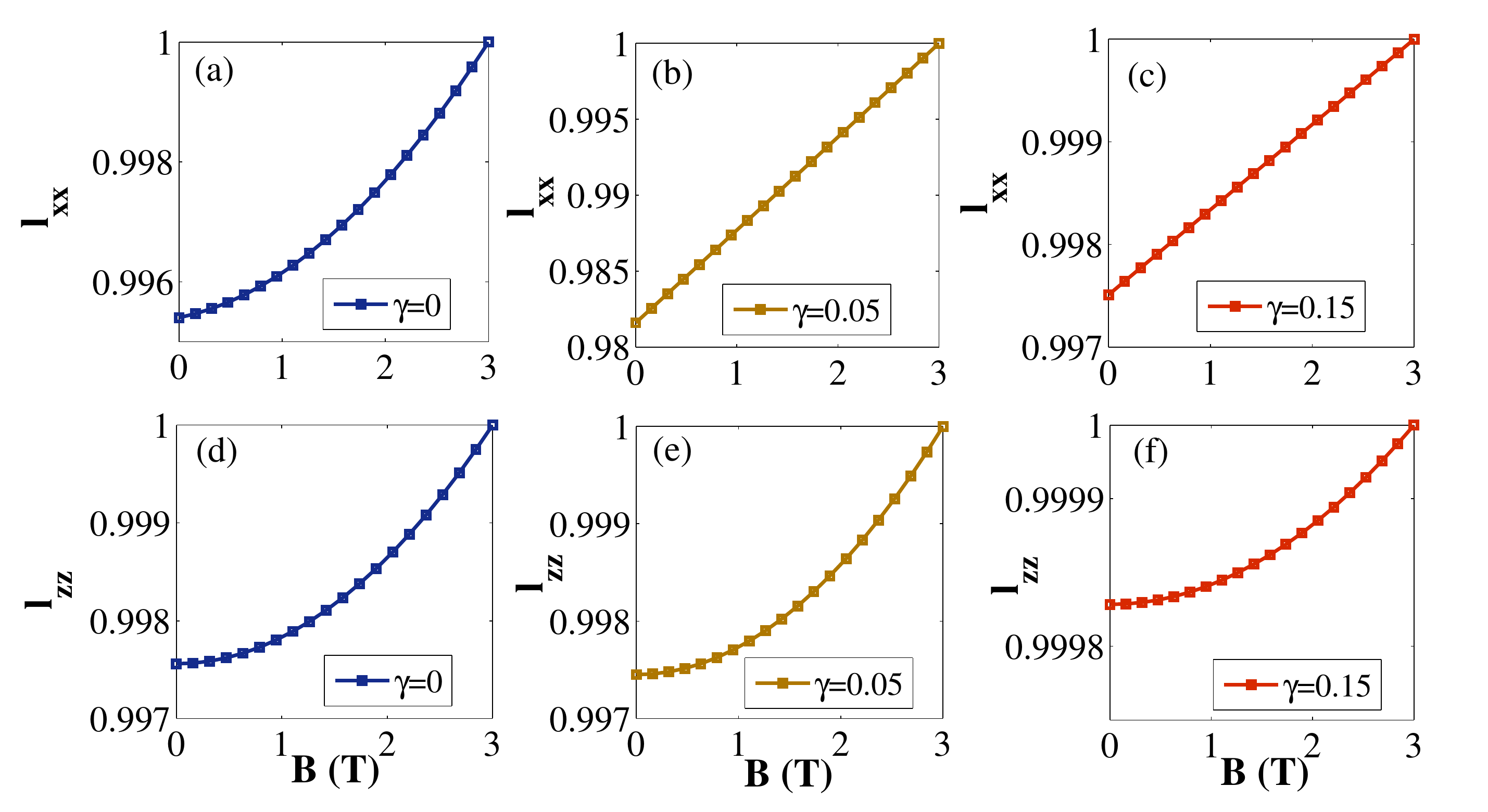,trim=0.10in 0.15in 0.0in 0.0in,clip=true, width=90mm}
\caption{(Color online)  First row (a)-(c): Longitudinal magneto-thermal conductivity $l_{xx}$ (normalized by $l_{xx}$ for $B=3$ T) calculated at $T=12$ K and {$\mu =0.04$ eV} as a function of magnetic field (B) applied along the tilt direction ($x$ axis), for a lattice model of WSM for (a) $\gamma=0$, (b) $\gamma=0.05$, and (c) $\gamma=0.15$ respectively. (d)-(f) depict the $l_{zz}$ (normalized) as a function of B applied perpendicular to the tilt direction (z axis) for the same set of parameters mentioned above.}
\label{l_latt}
\end{center}
\end{figure}

In this section, we compute the longitudinal magneto-thermal conductivity for a lattice model of type-I and type-II WSMs and discuss the B dependence and angular dependence of LMTC. The longitudinal magneto-thermal conductivity ($l_{xx}$ and $l_{zz}$) for lattice model of Weyl fermions is shown in Fig.~\ref{l_latt} for three different tilt parameters. The lattice model provides itself a physical ultra-violet energy cut-off to the low energy spectrum. 

\begin{figure}[htb]
\begin{center}
\epsfig{file=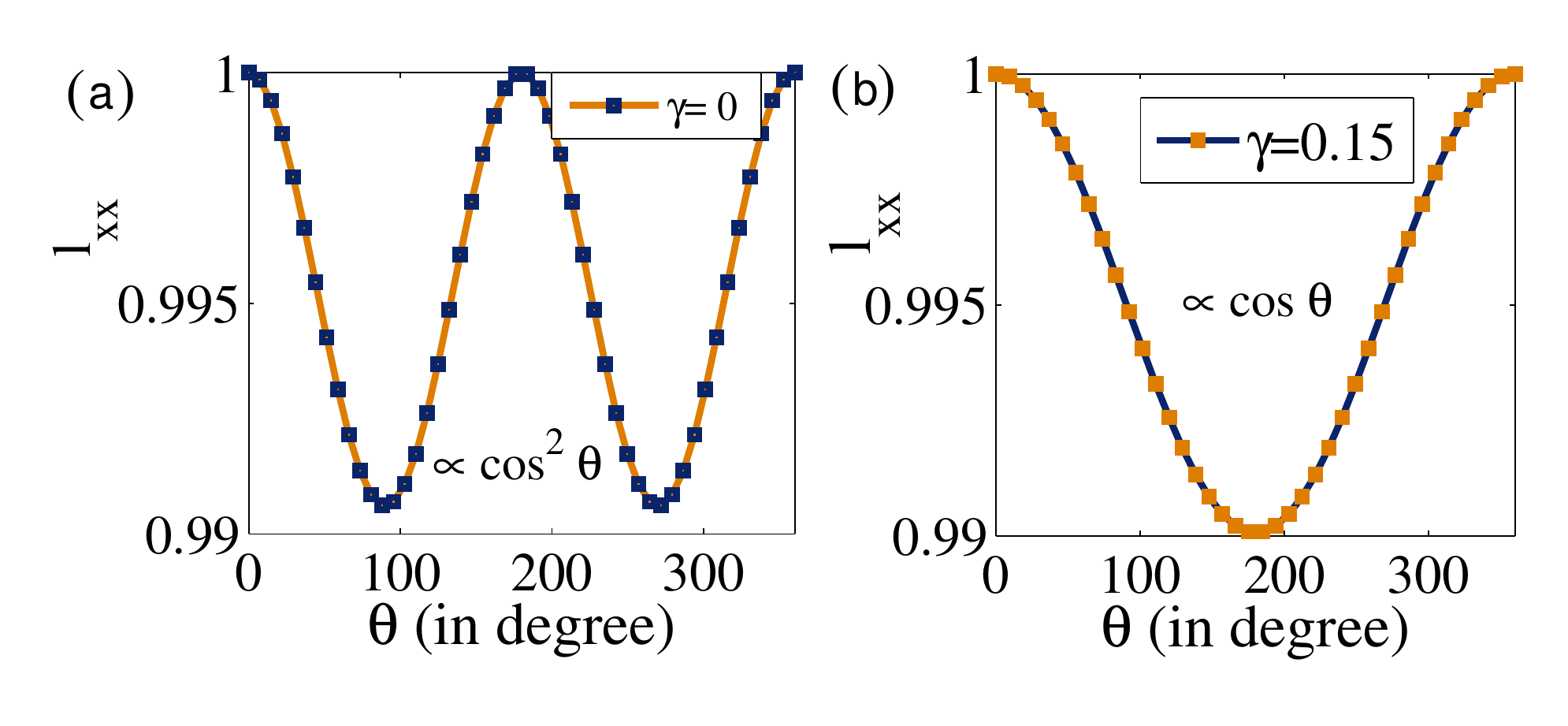,trim=0.10in 0.15in 0.0in 0.0in,clip=true, width=90mm}
\caption{(Color online)  (a)-(b) depict the angular dependence of longitudinal magneto-thermal conductivity at $B=3$T for type-I ($\gamma=0$) and type-II ($\gamma=0.15$) WSMs respectively. Here, we have taken $T=12$ K and $\mu=0.04$ eV. (We have normalized the $y$-axes of (a) and (b) by $l_{xx}(\theta=0)$)}
\label{Weyl_1}
\end{center}
\end{figure}

Fig.~\ref{l_latt} depicts $l_{xx}$ as a function of magnetic field at $T=12$ K for a TRS breaking type-I WSM ($\gamma=0$). In the absence of any tilt, LMTC follows quadratic B dependence as shown in figure.
Using Eq.~(\ref{e32}) we can now express $l_{xx}$ in terms of the diagonal components of the conductivity tensor, $l_{\parallel}$ and $l_{\perp}$, corresponding to the cases when the thermal current flows along and perpendicular to the magnetic field. Substituting $\theta=0$ and $\theta=\pi/2$ into Eq.~(\ref{e32}), we have
\begin{align}
&l^{\parallel}=l_{0}+e^{2}\int\frac{d^{3}k}{(2\pi)^{3}}\tau\frac{(\epsilon_{\mathbf{k}}-\mu)^{2}}{DT}\left(-\frac{\partial f_{0}}{\partial \epsilon}\right) \frac{B^2}{\hbar^2}(\mathbf{v_{k}}\cdot\mathbf{\Omega_{k}})^{2}, \nonumber \\
&l^{\perp}=l_{0}.
\label{s_para}
\end{align}
Eq.~(\ref{e32}) thus take the form
\begin{align}
l_{xx}=l^{\perp}+\Delta l \cos^{2} \theta,
\label{sxx_chiral}
\end{align}
where $\Delta l=l^{\parallel}-l^{\perp},$ gives the anisotropy in magneto-thermal conductivity due to chiral magnetic effect. The longitudinal magneto-thermal conductivity has the angular dependence of $\cos^{2} \theta$ which is shown in Fig.~\ref{Weyl_1}(a), leading to the anisotropic thermal resistance.
It is clear from the expression that LMTC has the finite contribution for all field directions.

In type-II WSM, we have calculated the LMTC for two different configurations; $l_{xx}$ appears in a configuration where both $\nabla T$ and B are parallel to tilt direction (i.e. along $x$ axis in present case) whereas $l_{zz}$ comes into play when both $\nabla T$ and B act perpendicular to tilt direction.

Fig.~\ref{l_latt} depicts $l_{xx}$ and $l_{zz}$ as a function of the magnetic field at $T=12$ K for a TRS breaking type-II WSM described by Eq.~(\ref{H_total}) for $\gamma=0.15$. Our calculation reveals that longitudinal magneto-thermal conductivity ($l_{xx}$) follows linear in B for the parallel set-up as shown in Fig.~\ref{l_latt}(c). On the other hand, $B^{2}$ dependence of LMTC has been found when B is applied perpendicular to the tilt direction ($l_{zz}$) as depicted in Fig.~\ref{l_latt}(f). It is clearly seen from the Eq.~(\ref{e32}) that both types of B dependence in LMTC arise due to chiral magnetic term. There is no qualitative difference in results for type-I and type-II WSM phases in the presence of anisotropy.

In the type-II WSM, the B-linear term in $l_{xx}$ becomes dominant due to the anisotropy which leads the computed longitudinal magneto-thermal conductivity to follow the $\cos \theta$ angular dependence at finite magnetic field for parallel set-up as shown in Fig.~\ref{Weyl_1}(b).


\section{Planar Thermal Hall Effect}
\label{Hall}

In this section, we discuss the numerical results of the novel effect PTHC for type-I and type-II WSMs. We have computed the B dependence and angular dependence of $l_{yx}^{\text{pth}}$ using Eq.~(\ref{e33}) for a TRS breaking WSM.

\begin{figure}[htb]
\begin{center}
\epsfig{file=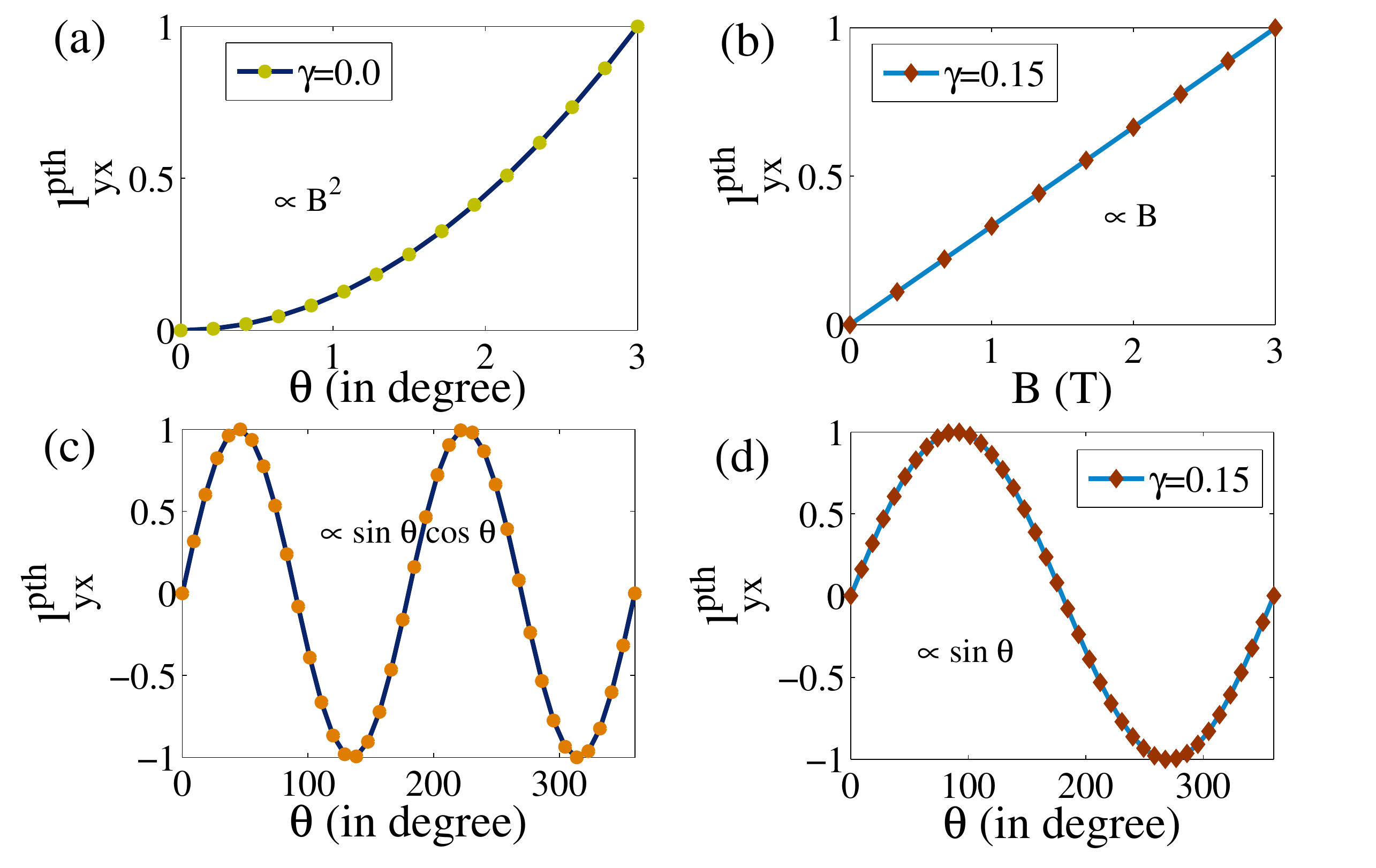,trim=0.0in 0.05in 0.0in 0.05in,clip=true, width=92mm}\vspace{0em}
\caption{(Color online) shows the normalized amplitude of planar thermal Hall conductivity at $\theta=\frac{\pi}{4}$ (normalized by $l_{yx}^{pth}$ at $B=3$ T) computed numerically for a Weyl semimetal with two Weyl nodes and tilt parameter (a) $\gamma=0$ (type-I WSM) and (b) $\gamma=0.15$ (type-II WSM) as a function of the magnetic field $\mathbf{B}$ applied along the tilt direction ($x$-axis). (c)-(d) depict the angular dependence of planar thermal Hall conductivity for $B=3$ T for type-I ($\gamma=0$) and type-II $(\gamma=0.15$) WSM phases respectively (B applied parallel to tilt direction). ($y$ axis is normalized by $l_{yx}^{pth}$ at $\theta=\frac{\pi}{4}$ and $\theta=\frac{\pi}{2}$ for figure (c) and (d) respectively.) {We have taken $\mu =0.04$ eV}.}
\label{WSM_2}
\end{center}
\end{figure}

We first examine the behavior of $l_{yx}^{\text{pth}}$ for the case $\gamma=0$, type-I WSM phase. Using Eqs.~(\ref{s_para}), we can now express $l_{yx}^{\text{pth}}$ as
\begin{align}
l_{yx}^{\text{pth}}=\Delta l \sin \theta \cos \theta.
\label{sxx_chiral}
\end{align}
The amplitude ($\Delta l$) of planar thermal Hall conductivity shows $B^{2}$-dependence for any angle except for $\theta=0$ and $\theta=\pi/2$ as shown in Fig.~\ref{WSM_2}(a). The planar thermal Hall conductivity follows the $\cos \theta \sin \theta$ dependence as depicted in Fig.~\ref{WSM_2}(c).

If we increase the $\gamma$ value then Weyl cones start to be tilted along the $k_{x}$ direction and the system stabilizes in type-II WSM phase after a critical value of $\gamma=0.1$. In Fig.~\ref{WSM_2}(b) we have plotted the numerically calculated PTHC ($l_{xy}^{pth}$ at $\theta=\pi/4$) for a type-II WSM as a function of $\mathbf{B}$. Our calculations reveal that the PTHC follows a B-linear dependence when B and $\nabla T$ are parallel to the tilt axis. For non-zero magnetic field, PTHC shows $\sin \theta$ dependence for the same configuration of the applied $\nabla T$ and $\mathbf{B}$ as shown in Fig.~\ref{WSM_2}(d). On the other hand, the B-dependence PTHC is quadratic when the $\nabla T$ and B are applied perpendicular to the tilt direction. In this configuration, PTHC follows the same angular dependence as in the case for type-I WSM with no tilt. We have also investigated the behavior of PTHC for type-I WSM with finite tilt. We find that PTHC shows similar angular and B dependence as in the case of type-II WSM.
\begin{figure}[htb]
\begin{center}
\epsfig{file=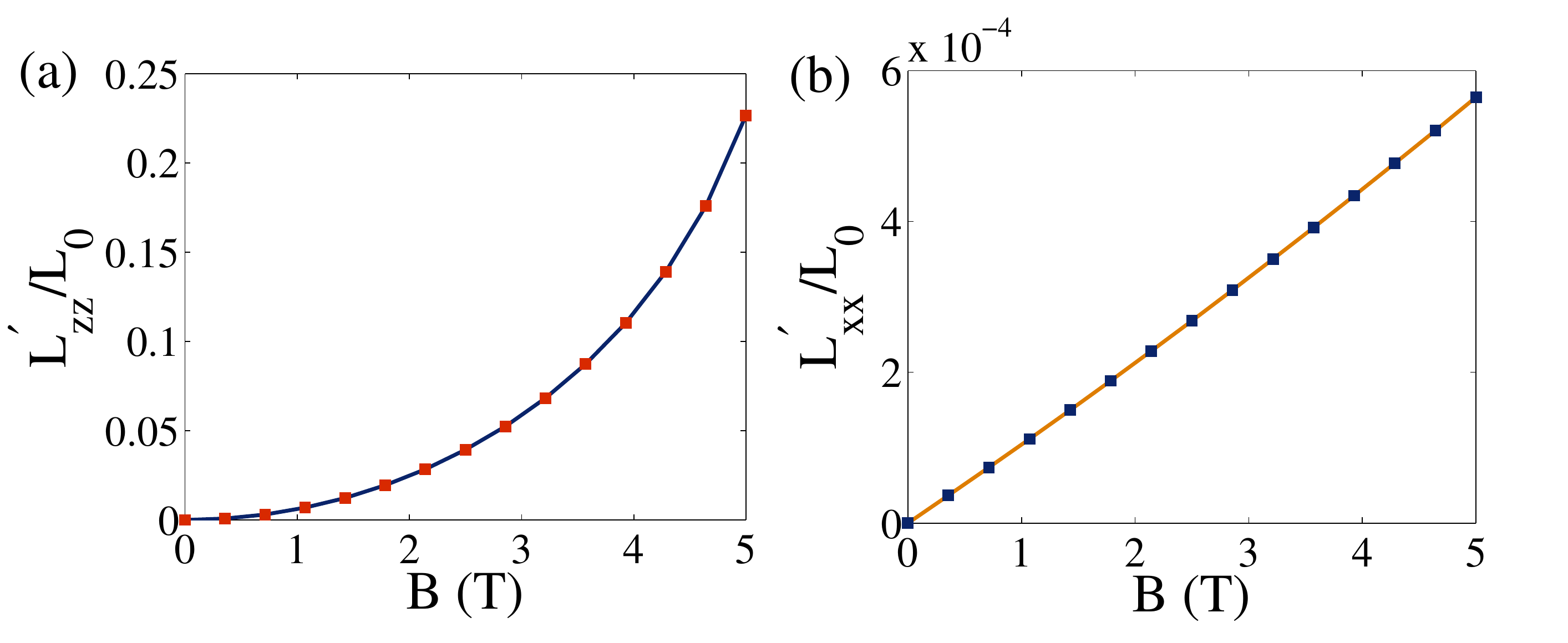,trim=0.10in 0.15in 0.0in 0.0in,clip=true, width=92mm}
\caption{(Color online)  Plot of $L^{\prime}/L_{0}=\frac{L(B)-L_{0}}{L_{0}}$ ($L_{0}$ is the Lorentz number) at $T=12$ K as a function of magnetic field (B) applied (a) perpendicular to the tilt direction ($z$ axis) and (b) along the tilt direction ($x$ axis) for an inversion broken Weyl semimetal WTe$_{2}$. The parameters are given in Table.~1. We have taken $\mu=0.065$ eV.}
\label{LT}
\end{center}
\end{figure}

\section{Wiedemann-Franz Law of an Inversion Symmetry Breaking Weyl Semimetal}
\label{wiedemann}
{The} Wiedemann-Franz law states that the ratio of electronic contribution of thermal conductivity and electrical conductivity for a metallic state is proportional to temperature. This law which holds for Landau Fermi Liquid, can be written as
\begin{eqnarray}
\frac{\kappa_{ij}}{\sigma_{ij}T}=L_{0},
\end{eqnarray}
where $L_{0}=\frac{\pi^{2}k_{B}^{2}}{3e^{2}}$ is the Lorentz number. 
{The Wiedemann-Franz law has been studied theoretically in the context of time reversal symmetry broken Weyl semimetals using various model Hamiltonians~\cite{Spivak_2016, kim_2014}. It has also been studied experimentally on the Dirac semimetal system Cd$_3$As$_2$  in the presence of a magnetic field~\cite{Mandal_2015}. Since all of these studies are on TRS broken Weyl semimetals, below we study the violation of the Wiedemann Franz law for an inversion broken WSM such as WTe$_{2}$.}
Recently, WTe$_{2}$ has been classified as an inversion broken type-II Weyl semimetal both theoretically~\cite{Soluyanov_2015} and experimentally~\cite{Wu_2016}. It has been found that WTe$_{2}$ contains 8 Weyl points in the $k_{z}$ = 0 plane and form a pair of quartets located at 0.052 eV and 0.058 eV above the Fermi level (E$_{F}$)~\cite{Soluyanov_2015}. Therefore, the linearized Hamiltonian for WTe$_{2}$ can be written as
\begin{align}
H(\textbf{k}) &= \Delta + Ak_{x} + Bk_{x} + (ak_{x} + ck_{y})\sigma_{y}+ (bk_{x} + dk_{y})\sigma_{z} \nonumber \\
&+fk_{z}\sigma_{x}.
\end{align}
The parameter values for WTe$_{2}$ obtainied by fitting the Hamiltonian to the \textit{ab inito} band structure calculation~\cite{Soluyanov_2015}, are given in Table I.
\begin{table}[h]
\centering
\caption{First-principles fitted parameter values in eV-\text{\r{A}} as given in Reference~\cite{Soluyanov_2015}. Here, the energy of the WPs corresponds to the energy above E$_{\text{F}}$. }
\label{tab:Table1}
\begin{tabular}{c c c c c c c c}
\hline
\hline
    \textbf{Energy of the WPs} & \textbf{$A$} &  \textbf{$B$} & \textbf{$a$} & \textbf{$b$} & \textbf{$c$} & \textbf{$d$} & \textbf{$f$} \\
    \hline
    \hline
    0.052 eV & -2.739 & 0.612 & 0.987 & 1.107 & 0.0 & 0.270 & 0.184 \\
    0.058 eV & 1.204 & 0.686 & -1.159 & 1.046 & 0.0 & 0.055 & 0.237 \\
    \hline
    \hline
\end{tabular}
\end{table}


We have computed both longitudinal thermal conductivity and longitudinal electrical conductivity for WTe$_{2}$. {The longitudinal electrical conductivity ($\sigma_{jj}$) is given by 
\begin{eqnarray}
\sigma_{jj}=e^{2}\int\frac{d^{3}k}{(2\pi)^{3}}\tau D^{-1}[({v_{j}}+
\frac{eB_j}{\hbar}(\mathbf{v_{k}}\cdot\mathbf{\Omega_{k}}))^{2} 
]\left(-\frac{\partial f_{0}}{\partial \epsilon}\right). \nonumber \\
\label{eq_cm1}
\end{eqnarray}}

{In order to calculate the L$_{zz}$ and L$_{xx}$, we set the chemical potential 0.065 eV which is just above the second Weyl node (0.058 eV). First we calculate these quantities for each node separately and then add the contributions for the different nodes.} Interestingly, it turns out that the Wiedemann-Franz law is violated and becomes B dependent for this material due to both the chiral magnetic effect and chiral anomaly. In Fig.~\ref{LT} we have plotted the deviation of Lorentz number from its standard value $L_{0}$ ($L^{\prime}/L_{0}=\frac{L(B)-L_{0}}{L_{0}}$) as a function of applied magnetic field. Our calculation reveals that the deviation of Lorentz number ($L^{\prime}$) follows quadratic B dependence when the external fields are applied perpendicular to the tilt direction ($z$ axis) as shown in Fig.~\ref{LT}(a). On the other hand, linear B-dependence of $L^{\prime}$ has been found when applied fields are parallel to the tilt direction ($x$ axis) as depicted in Fig.~\ref{LT}(b). In both cases, the sign of $L^{\prime}$ becomes positive which indicates that the ratio of thermal to electrical conductivity will increase from its standard value with the applied field. {Our results on the violation of the Wiedemann-Franz law in WTe$_2$, which is an inversion broken WSM, agree with the previous studies where the case of time reversal broken WSM was considered both theoretically~\cite{kim_2014} and in experiments~\cite{Mandal_2015}.} 
\section{Conclusions}
\label{summary}

We present a quasi-classical theory of chiral magnetic effect induced planar thermal Hall effect in Weyl semimetals. We show that when the thermal gradiant and magnetic field are applied in-plane but not aligned parallel to each other, a non-zero planar thermal Hall response arises strictly out of the chiral magnetic effect. This Hall effect is of a different nature from the usual Lorentz force mediated thermal Hall response and even the Berry phase mediated anomalous thermal Hall response. We derive an analytical expression for planar thermal Hall conductivity and investigate its generic behavior for type-I and type-II WSMs. Interestingly, we find that PTHC follows the $B^{2}$ dependence in type-I Weyl semimetal (anisotropy parameter $\gamma=0$, see Eq.~\ref{H_tilt}) whereas it is linear in B in type-II Weyl semimetal when B and $\nabla T$ are applied along the tilt direction. The angular dependence of PTHC also changes from $\sim \cos \theta \sin \theta$ to $ \sim \sin \theta$ as we go from type-I WSM ($\gamma=0$) to type-II WSM. In type-II WSM, when both B and $\nabla T$ are applied perpendicular to the tilt direction, the PTHC shows the conventional B$^{2}$-dependence as in the case of type-I WSM ($\gamma=0$). {Although the behavior of planar thermal Hall conductivity and longitudinal magneto-thermal conductivity are similar to the behavior of their electric field counterparts~\cite{Nandy_2017} it is important to emphasize that the origin of PTHC is chiral magnetic effect (third term on the right hand side of Eq.~(\ref{e9})) while the planar Hall effect in Ref.~47 is due to the chiral anomaly term in Eq.~(\ref{e10}) (third term on the right hand side of Eq.~(\ref{e10})). The measurement of PTHC also involves a different experimental geometry than that of PHE. The current work is expected to stimulate experimental efforts to uncover PTHC which can be taken as an experimental signature of chiral magnetic effect in WSMs.}

Additionally, we also investigate the longitudinal magneto thermal conductivity in Weyl semimetals and the   violation of the Wiedemann-Franz law in inversion broken type-II Weyl semimetal such as WTe$_{2}$. We find that Wiedemann-Franz Law is violated in WSMs due to both chiral magnetic effect and chiral anomaly.

\section{Acknowledgements}
SN and AT acknowledge the computing facility from DST-Fund for S and T infrastructure (phase-II) Project installed in the Department of Physics, IIT Kharagpur, India. SN  acknowledges  MHRD, India for support. ST acknowledges support from ARO Grant No: (W911NF-16-1-0182).

\end{document}